\documentclass{ws-ijmpd}

\begin{document}

\markboth{Feoli and Mancini} {A fundamental equation for
Supermassive Black Holes}

%
\catchline{}{}{}{}{}
%

\title{A fundamental equation for Supermassive Black Holes}

\author{ANTONIO FEOLI}

\address{Department of Engineering, University of Sannio\\
Corso Garibaldi n. 107, Palazzo Bosco Lucarelli \\ %
82100 - Benevento, Italy\\
feoli@unisannio.it}

\author{LUIGI MANCINI}

\address{Max Planck Institute for Astronomy\\
K\"{o}nigstuhl 17 \\
69117 - Heidelberg, Germany \\
mancini@mpia-hd.mpg.de}

\maketitle

\begin{history}
\end{history}

\begin{abstract}
We developed a theoretical model able to give a common origin to
the correlations between the mass $M_{\bullet}$ of supermassive
black holes and the mass, velocity dispersion, kinetic energy and
momentum parameter of the corresponding host galaxies. Our model
is essentially based on the transformation of the angular momentum
of the interstellar material, which falls into the black hole,
into the angular momentum of the radiation emitted in this
process. In this framework, we predict the existence of a relation
of the form $M_{\bullet} \propto R_{\mathrm{e}} \sigma^3$, which
is confirmed by the experimental data and can be the starting
point to understand the other popular scaling laws too.
\end{abstract}


\section{Introduction}
At present, thanks to the improved angular resolution of modern
telescopes, the experimental evidence indicates that a variety of
nearby galaxies host a supermassive black hole (SMBH;
$M_{\bullet}>10^6 M_{\odot}$) at their
center\cite{kormendy95}\cdash\cite{richstone98}. The subsequent
discovery of a large number of scaling laws, in which the mass of
SMBHs correlates with the properties of the host galaxies
(bulges)\cite{magorrian98}\cdash\cite{soker10}, demonstrates a
link between the process of accretion of SMBHs and the formation
and evolution of their galaxy.

Even if many analytical and numerical models have been proposed at
the same time in order to explain the observed scaling
relationships (see for example
Refs.~\refcite{churazov02}--\refcite{lusso10}), the physical
origin of these correlations, as well as the answer to the
question ``what is the most fundamental one?'', are still unclear
and under debate\cite{novak06}\cdash\cite{lauer07}.

Here we propose a simple theoretical model able to give a common
origin for the correlations between the mass $M_{\bullet}$ of
SMBHs and the mass, velocity dispersion, kinetic energy and
momentum parameter of the corresponding bulges. Starting from a
principle of conservation of the angular momentum and using, as a
first approximation, the Bondi--Hoyle--Lyttleton (BHL) theory of
accretion\cite{hoyle39}\cdash\cite{horedt00}, we found a
fundamental equation of the form $M_{\bullet} \propto
R_{\mathrm{e}} \sigma^3$, where $R_{\mathrm{e}}$ and $\sigma$ are
the bulge effective radius and effective stellar velocity
dispersion, respectively. From the projections of this fundamental
plane, using suitable correlations, we easily derive the other
popular scaling laws. Despite the drastic hypotheses and hence the
simplicity of the model, our results show an excellent agreement
between its predictions and the experimental data, indicating that
we have found a basic law for galaxies and their SMBHs.
%

\section{The model}
Marconi and Hunt in 2003 were the first to note that $M_{\bullet}$
is significantly correlated both with $\sigma$, and with
$R_{\mathrm{e}}$. The conclusion of their study was that a
combination of $\sigma$ and $R_{\mathrm{e}}$ is necessary to drive
the correlations between $M_{\bullet}$ and other bulge
properties\cite{marconi03}. This topic was then deeply
investigated through simulations of major galaxy mergers, which
defined a fundamental plane, analogous to the fundamental plane of
elliptical galaxies, of the form $M_{\bullet} \propto
R_{\mathrm{e}}^{1/2} \sigma^3$ or $ M_{\bullet} \propto
M^{1/2}_{\star} \sigma^2$, and $M_{\bullet} \propto (M_{\star}
\sigma^2)^{0.7}$, where $M_{\star}$ is the bulge stellar
mass\cite{hopkins07a}\cdash\cite{marulli08}. These scaling laws
are very similar to what was really found
observationally\cite{hopkins07b}\cdash\cite{feoli10}. Following
this path we build a theoretical model able to explain  all the
famous relations linking the mass of SMBHs with the properties of
their bulges.

Our model is mainly based on two hypotheses: the conservation of
angular momentum and a suitable velocity field of the gas. Then we
need to make an approximation to estimate the accretion radius of
the black hole, which can be found either in a rough way or
recurring to the BHL theory of accretion.

Let us consider a black hole of mass $M_{\bullet}$ emitting
radiation at rate $L_\varepsilon$, and accreting from a
distribution $\rho$ of gas, whose inward velocity is
$V_{\mathrm{in}}$.

 First, we assume that the
angular momentum is conserved in such a way that a part
($M_{\mathrm{acc}}$) of the total mass of the gas contained in the
galactic bulge will be captured by the black hole, converting its
angular momentum  into the angular momentum of the perpendicularly emitted
radiation (with an effective mass $M_{\mathrm{rad}}$). Of course,
there are other mechanisms of conversion, transport or dissipation
of angular momentum (viscosity, change in the spin of central
black hole, etc.) that, in the past, have contributed with a
different weight and importance to the accretion of black holes
and probably are still acting now, but our drastic hypothesis
focuses the attention only on the above described process. More
complicated and detailed models should involve effects due to the
other mechanisms that in our approximation are neglected.

An order of magnitude estimate of the angular momentum for the
galactic bulge\cite{emsellem07}\cdash\cite{jesseit09} leads to the
conservation equation:
%
\begin{equation}
M_{\mathrm{acc}} \, R_{\mathrm{e}} \, V_{\mathrm{rot}}=c \,
M_{\mathrm{rad}} \, R_{\mathrm{A}},%
\label{Eq_01}
\end{equation}
where $R_{\mathrm{e}}$ is the effective radius of the bulge,
$V_{\mathrm{rot}}$ is the mass-weighted mean rotational velocity
of the gas, $c$ is the speed of light and $R_{\mathrm{A}}$ the
accretion radius of the Black hole.

Then, we suppose that the velocity of the gas in the galactic
bulge is related to the effective   stellar velocity dispersion $\sigma$ in
such a way that $V_{\mathrm{in}}= \sigma$ (see
Refs.~\refcite{soker10} and \refcite{soker09}) and
$V_{\mathrm{rot}}= A \, \sigma$, where $A$ is a
constant\cite{baes03}. In the ideal case of the isothermal sphere,
$A$ is equal to $\sqrt{2}$.

Now we must estimate the accretion radius and as a first
approximation we can use the Bondi--Hoyle--Lyttleton
theory\cite{hoyle39,bondi44}. In their model the rate at which the
gas is accreted onto the black hole is
%
\begin{equation}
\dot{M}_{\mathrm{acc}}=\pi R_{\mathrm{B}}^2 V_{\mathrm{in}}
\rho, %
\label{Eq_02}
\end{equation}
where
%
\begin{equation}
R_{\mathrm{B}}=2GM_{\bullet}/V_{\mathrm{in}}^2%
\label{Eq_03}
\end{equation}
is the BHL radius and $G$ is the gravitational constant. Of course
the BHL theory is a realistic model in the case of radial
accretion with low transverse velocity but it can be used as a
good approximation even in presence of  angular momentum (see the
Appendix A). Anyway, also without considering the BHL theory, the
expression of $R_{\mathrm{B}}$ in Eq. (3), with $V_{\mathrm{in}}
\simeq \sigma$, roughly corresponds to the well known
gravitational radius of influence of a black hole, which we can
adopt as an estimate of $R_{\mathrm{A}}$.

Hence, we consider $R_{\mathrm{A}} \simeq R_{\mathrm{B}}$ (see the Appendix A), and
substitute Eq. (3) in Eq. (1). Finally, recalling our hypothesis
on the velocity field, we obtain the fundamental equation for
supermassive black holes we were seeking:
%
\begin{equation}
M_{\bullet} =  \frac{A \, R_{\mathrm{e}} \sigma^3}{2 \, \varepsilon \, G \, c }
\simeq 4.4 \times 10^7 \left(\frac{A}{\sqrt{2}} \right) %
\left(\frac{0.1}{\varepsilon} \right) \left(\frac{R_{\mathrm{e}}}{\mathrm{kpc}}\right) \left(\frac{\sigma}%
{200 \, \mathrm{km/s}}\right)^3,%
\label{Eq_04}
\end{equation}
where $\varepsilon$ is the mass to energy conversion efficiency,
which is set by the amount of rest mass energy of matter accreted
onto the black hole that is extracted and radiated outward
($\varepsilon=M_{\mathrm{rad}}/M_{\mathrm{acc}}$).

The accreting black hole liberates energy at a rate
%
\begin{equation}
L_{\varepsilon} = \varepsilon  \dot{M}_{\mathrm{acc}} c^2 = \ell
L_{\mathrm{Edd}}. %
\label{Eq_05}
\end{equation}
It is commonly assumed that the accretion of matter onto a black
hole releases energy at $10\%$ efficiency so that a fixed value of
$\varepsilon = 0.1$ is usually adopted\cite{shakura73}, even if
the range $0.001 \leq \varepsilon \leq 0.1$ has been recently
investigated\cite{lusso10}.

This radiated luminosity is related to the Eddington limit,
$L_{\mathrm{Edd}}$. In particular, for $\ell = 1$ the central
black hole is radiating at its Eddington limit:
%
\begin{equation}
L_{\mathrm{Edd}} = \frac{4 \pi G M c \,
m_{\mathrm{p}}}{\sigma_{\mathrm{T}}} = \frac{4 \pi G M
c}{\kappa_{\mathrm{Edd}}}. %
\label{Eq_06}
\end{equation}
Here $m_\mathrm{p}$ is the proton mass, $\sigma_{\mathrm{T}}$ the
Thomson scattering cross--section of the electron, and
$\kappa_{\mathrm{Edd}} = \sigma_{\mathrm{T}}/m_{\mathrm{p}}$ the
opacity of the fully ionized hydrogen. Substituting Eq.
(\ref{Eq_02}) and Eq. (\ref{Eq_06}) in Eq. (\ref{Eq_05}), and
recalling our two hypotheses, we can  determine also the gas
density at the effective radius in a form:
%
\begin{equation}
\rho = \frac{2 \, \ell}{\kappa_{\mathrm{Edd}} \, A \,
R_{\mathrm{e}}}, %
\label{Eq_07}
\end{equation}
similar to that found in Ref.~\refcite{begelman05}. While in other models
the gas density is given among the hypotheses, in our approach it is a consequence of the
theory.

We have tested the effectiveness of our model on a
sample\cite{hu09} of 58 nearby galaxies ($z\sim0$). By using these
experimental data and the corresponding errors, we report in Table
1 the best--fitting values for the slope $m$ and the normalization
$b$ for the linear relations used in this work. These values have
been calculated by the routine FITEXY\cite{press92} for the
relation $y=b+mx$, by minimizing the $\chi^2$. The  estimates of
the $\chi_{\mathrm{red}}^2=\chi^2/(58-2)$ and the Pearson linear
correlation $r$ for each relation are also shown. Errors have been
calculated by using the formula reported in Appendix A of
Ref.~\refcite{feoli09}. The results of the fits for the
$M_{\bullet} - M_{\mathrm{G}}$ and $M_{\bullet} - M_{\mathrm{G}}
\sigma^2$ relations are in agreement with the ones obtained in a
previous paper using three different samples of data\cite{feoli10}
and with the results obtained by other authors while the
$M_{\bullet} - \sigma$
relation requires a more careful discussion (see Appendix B). %
%
\begin{table}[ph]
\tbl{Black hole--bulge correlations and fitting parameters for the
considered galaxy sample.}%
{\begin{tabular}{@{}cccccc@{}}\toprule Relation & $b\pm\Delta b$ &
$m\pm\Delta m$ & $\chi_{\mathrm{red}}^2$ & $\epsilon_{0}$& $r$\\
\colrule
  $M_{\bullet}-R_{\mathrm{e}}\sigma_{200}^3$ & $8.11\pm 0.03$ & $1.00\pm 0.02$ & $5.00$ & $0.39$ & $0.87$ \\
  $M_{\bullet}-M_{\mathrm{G}}\sigma^2/c^2$ & $4.56\pm 0.10$ & $0.87\pm 0.02$ & $6.03$ & $0.40$ & $0.87$ \\
  $M_{\bullet}-\sigma_{200}$ & $8.21\pm 0.02$ & $5.83\pm 0.15$ & $6.22$ & $0.40$ & $0.87$ \\
  $M_{\bullet}-M_{\mathrm{G}}\sigma/c$ & $0.73\pm 0.19$ & $1.01\pm 0.03$ & $6.61$ & $0.42$ & $0.87$ \\
  $M_{\bullet}-M_{\mathrm{G}}$ & $8.74\pm 0.03$ & $1.21\pm 0.03$ & $7.71$ & $0.45$ & $0.85$ \\
\colrule
  $R_{\mathrm{e}}-\sigma_{200}$ & $0.12\pm 0.01$ & $2.72\pm 0.09$ & $10.31$ & $0.32$ & $0.75$ \\
  $R_{\mathrm{e}}-M_{\mathrm{G}}\sigma^2/c^2$ & $-1.77\pm 0.06$ & $0.45\pm 0.01$ & $3.26$ & $0.16$ & $0.92$\\   \botrule
\end{tabular} \label{ta1}}
\end{table}
The best--fitting line for the new relationship
$M_{\bullet}-R_{\mathrm{e}} \sigma_{200}^3$ in a log--log plane
is:
%
\begin{equation}
\log_{10} M_{\bullet} = (8.11 \pm 0.03)+ (1.00 \pm 0.02)\log_{10}
(R_{\mathrm{e}} \sigma_{200}^3),%
\label{Eq_08}
\end{equation}
where $ \sigma_{200}$ is the velocity dispersion in $200$ km
sec$^{-1}$ units, while $R_{\mathrm{e}}$ is expressed in kpc. The
linear relation (\ref{Eq_08}) has a slope equal to the unity,
which is exactly the value predicted by our model in the Eq.
(\ref{Eq_04}).

We remark also that this relation has the best $\chi^2$  and $r$
among the relations in the upper part of Table 1 involving the
black hole mass . We also report  the intrinsic
scatter $\epsilon_{0}$, finding that the relation
$M_{\bullet}-R_{\mathrm{e}}\sigma^3$ has $\epsilon_{0}=0.39$,
whereas the other ones have values of $\epsilon_{0} \geq 0.40$.

Furthermore, from the normalization we can calculate the effective
efficiency coefficient, which turns out to be equal to
$\hat{\varepsilon}= 2 \varepsilon/A = 0.048\pm 0.003$, a value
close to $0.05$ estimated in Ref.~\refcite{churazov02}, and to
$0.06$ used in the case of Schwarzschild's metric\cite{shakura73}.
Of course if the coefficient $\hat{\varepsilon}$ depends on $R_e$
or $\sigma$ the relation (8) will still exist but its slope will be different from the unity.

In Figure 1, we report the $M_{\bullet}-R_{\mathrm{e}}
\sigma_{200}^3$ diagram in a log–-log plot (we associated a
particular symbol to each galaxy according to its
morphology\cite{hu09}). The best-fitting line is also shown.

We also tested Eq. (\ref{Eq_04}) by using an old sample of 37 objects\cite{marconi03},
 obtaining a slope equal to $0.90\pm0.04$,
which is in good agreement with the slope of Eq. (8). Using a subset of 27
galaxies of the same old sample, Hopkins et al.\cite{hopkins07b} obtained a
fundamental plane $M_{\bullet} \propto R_{\mathrm{e}}^{0.43}
\sigma^{3.00}$ and Graham\cite{graham08} $M_{\bullet} \propto
R_{\mathrm{e}}^{0.28} \sigma^{3.65}$. The difference with our
result is due to a different fitting method and to the fact that
these authors have considered three free parameters. With a
different sample also Aller and Richstone\cite{aller07} studied
the same three parameters fit obtaining $M_{\bullet} \propto
R_{\mathrm{e}}^{0.28} \sigma^{3.16}$.

The fact that we predict a relation of the kind  $M_{\bullet}
\propto R_{\mathrm{e}} \sigma^{3}$ and not the Hopkins
result\cite{hopkins07a} $M_{\bullet} \propto R_{\mathrm{e}}^{0.5}
\sigma^{3}$ does not depend on an oversimplification of our model.
To obtain the latter relation one must start from  different hypotheses
and only increasing the number of experimental data it will be
possible to distinguish  the better approach.

\begin{figure}[pb]
\centerline{\psfig{file=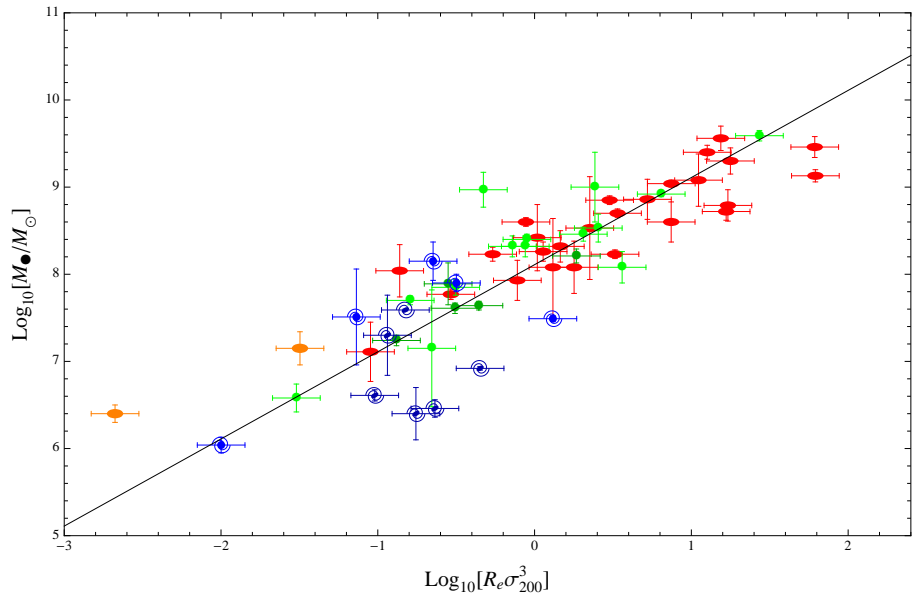,width=12.cm}} %
\vspace*{8pt} \caption{The
$M_{\bullet}-R_{\mathrm{e}}\sigma_{200}^3$ relation for the
galaxies of the considered data set, where $\sigma_{200}$ is the
bulge velocity dispersion in units of 200 km s$^{-1}$. The symbols
represent elliptical galaxies (red ellipses), lenticular galaxies
(green circles), barred lenticular galaxies (dark green circles),
spiral galaxy (blue spirals), barred spiral galaxies (dark blue
barred spirals), and dwarf elliptical galaxies (orange ellipses).
The black line is the line of best fit for the sample of galaxies
considered.}
\end{figure}

\section{The origin of the other scaling relations}
The dynamical masses of bulges, $M_{\mathrm{G}}$, can be estimated
by
%
\begin{equation}
M_{\mathrm{G}}= \frac{k R_{\mathrm{e}}\sigma^2}{G},%
\label{Eq_09}
\end{equation}
where $k$ is a model dependent dimensionless constant. We stress
that the isothermal model (i.e. $\sigma$ is a constant throughout
the galaxy) is not a hypothesis of our framework. We make use of
this approximation in order to test our model. As in
Refs.~\refcite{marconi03} and \refcite{hu09}, we use $k=3$ to
compute the ``isothermal masses'' of the galaxy sample
considered\cite{hu09} (in a more detailed model $k$ is dependent
on the Sersic index). Thanks to the fundamental Eq. (\ref{Eq_04}),
coupled with Eq. (\ref{Eq_09}), in the following we derive all the
most famous scaling relationships between the black hole mass and
the parameters of the host galaxy.

\subsection{The $M_{\bullet}-M_{\mathrm{G}} \, \sigma/c$ relation}
Replacing Eq. (\ref{Eq_09}) in Eq. (\ref{Eq_04}), we get:
%
\begin{equation}
M_{\bullet} = \frac{1}{\hat{\varepsilon} k} \left(\frac{M_{\mathrm{G}} \sigma}{c}\right),%
\label{Eq_10}
\end{equation}
in optimum agreement with the corresponding relation in Table 1,
where the value of the slope $1.01$ is close to unity as we
expected (the quantity $M_{\mathrm{G}} \sigma/c$ is also known as
the momentum parameter\cite{soker10}). If we impose the exponent
of the momentum to be equal to 1.00, then by refitting the data we
obtain a normalization $b=0.82 \pm 0.02$, from which we derive
$\hat{\varepsilon}= 0.05$.
%
\subsection{The $M_{\bullet} - \sigma$ relation}
Substituting the experimental relation between $R_{\mathrm{e}}$
and $\sigma$ (see Table 1) in Eq. (\ref{Eq_04}), we obtain:
%
\begin{equation}
M_{\bullet} = \frac{10^{0.12}}{\hat{\varepsilon} c \, G}
\left(\sigma_{200}\right)^{5.72}= 10^{8.23}\left(\sigma_{200}\right)^{5.72},%
\label{Eq_11}
\end{equation}
which is in agreement, inside the errors, with the corresponding
law in Table 1:
%
\begin{equation}
M_{\bullet} =10^{8.21 \pm 0.02}(\sigma_{200})^{5.83 \pm
0.15}.%
\label{Eq_12}
\end{equation}

\subsection{The $M_{\bullet} - M_{\mathrm{G}}\sigma^2/c^2$ relation}
Using Eq. (\ref{Eq_09}), Eq. (\ref{Eq_04}) can be written in terms
of the kinetic energy of random motions:
%
\begin{equation}
M_{\bullet} = \frac{1}{\hat{\varepsilon}} \left(\frac{c^2
R_{\mathrm{e}}}{G k^3}\right)^{1/4} \left(
\frac{M_{\mathrm{G}} \sigma^2}{c^2}\right)^{3/4}.%
\label{Eq_13}
\end{equation}
Replacing $R_{\mathrm{e}}$ in Eq. (\ref{Eq_13}) with the value
$R_{\mathrm{e}} = 10^{-1.77} (M_{\mathrm{G}}\sigma^2/c^2)^{0.45}$,
taken from the experimental relation in Table 1, we obtain:
%
\begin{equation}
M_{\bullet} =
10^{4.60}(M_{\mathrm{G}} \sigma^2/c^2)^{0.86},%
\label{Eq_14}
\end{equation}
where we have used the value of $0.048$ for $\hat{\varepsilon}$.
We point out the good match between the relation in Eq.
(\ref{Eq_14}) and the corresponding one reported in Table 1:
%
\begin{equation}
M_{\bullet} = 10^{4.56 \pm 0.10}(M_{\mathrm{G}}
\sigma^2/c^2)^{0.87 \pm 0.02}.%
\label{Eq_15}
\end{equation}
As shown in previous papers\cite{feoli09,feoli10}, this relation
has the best $\chi^2$ compared to $M_{\bullet}-\sigma$ and
$M_{\bullet}-M_{\mathrm{G}}$ laws.

\subsection{The $M_{\bullet} - M_\mathrm{G}$ relation}
Starting from the Eq. (\ref{Eq_10}) and expressing $\sigma$ in
terms of $M_{\bullet}$ (see Table 1), we obtain:
%
\begin{equation}
M_{\bullet} =
10^{-4.52} M_{\mathrm{G}}^{1.21},%
\label{Eq_16}
\end{equation}
in excellent agreement with the relation:
%
\begin{equation}
M_{\bullet} = 10^{-4.57 \pm 0.34} M_{\mathrm{G}}^{1.21 \pm
0.03}.%
\label{Eq_17}
\end{equation}
%

\section{Summary}
To sum up, we have shown that the model proposed in \S 2 works
very well since, given a consistent set of data, it perfectly
predicts the slope of a new relation (Eq. \ref{Eq_08}). Moreover,
the most investigated scaling relationships can be easily obtained
as projections of the plane identified by the fundamental Eq.
(\ref{Eq_04}). Unfortunately, we cannot infer this correlation at
redshift $z>>0$, because we are limited by the small number of
observable hosts. Other mechanisms of accretion, different from
the one presented in this work, may have acted in the past, and
hence other fundamental equations may have ruled the first
Gigayears of the life of SMBHs. Future detection of new SMBHs,
especially at higher redshift, and measurements of their masses
will enable us to confirm the universality of our law or if it
holds just for a restricted period of the cosmic time. At this
stage it is early to predict the consequences of our equation in
the context of the models about the co--evolution of galaxies and
black holes. First of all we must check the validity of the
approach using an enlarged sample of data. Recently Sani et al.
collected a new interesting (but not larger) galaxy
sample\cite{sani11} and we intend to present a complete analysis
of their data in a forthcoming paper. A preliminary result is that
a tight relation $M_{\bullet} - R_\mathrm{\mathrm{e}} \sigma^3$
exists, but its slope oscillates from 0.78 to 1.2 depending on the
number of pseudobulges\footnote{Essentially, a pseudobulge is a
bulge that shows photometric and kinematic evidence for disk--like
dynamics} considered in the fit. This aspect requires a deeper
investigation before drawing any conclusions about the evolution
of supermassive black holes and galaxies.

\section*{Acknowledgments}
We are grateful to Gaetano Scarpetta, and
Sidney van den Bergh for their very useful suggestions and to the anonymous
referee whose comments have contributed to improve our paper. L.M.
acknowledges support for this work by research funds of the
University of Sannio, University of Salerno, and the International
Institute for Advanced Scientific Studies. A.F. acknowledges
support for this work by research funds of the University of
Sannio.

\section*{Appendix A}
We underline that our model is not based on the BHL theory
assumption. We can simply estimate the value of the accretion
radius recurring to the concept of radius of influence of a black
hole. Actually in our paper the BHL theory is equivalently used as
another approach to have an estimate of the accretion radius even
if it is well known that in presence of angular momentum the BHL
approach is only a fair approximation.

In their original paper, Hoyle and Lyttleton suppose that an
element of volume of the gas cloud has an initial angular
momentum, but it ``\textit{loses this momentum through its
constituent particles suffering collisions}'' at a radius
$R_{\mathrm{B}}$\cite{hoyle39}. That is, the transverse velocities
of the particles, which reach the accretion line from opposite
directions, annihilate reciprocally, whereas the radial component,
if it is not bigger than the escape velocity (and this occurs at a
distance $ d \leq R_{\mathrm{B}}$ from the hole), makes possible
that the particles were captured by the black hole. Then they show
that the collisions occur with sufficient frequency to be
effective in reducing the angular momentum.

Anyway, it is possible to take into account the angular momentum,
but the final result does not drastically change. In fact, in a
generalization of BHL approach, the author in
Ref.~\refcite{horedt00} considers the orbital velocity as the
source of a lateral pressure on the accretion column. Having
included this pressure term in the dynamical equations, he
obtained an accretion radius $R_{\mathrm{A}}$ of the same order of
magnitude of the BHL radius: $0.3 \, R_{\mathrm{B}} <
R_{\mathrm{A}} < 1.75 \, R_{\mathrm{B}}$. Both BHL and Horedt
analyzed a case of pure accretion of the central object without
considering the energy eventually radiated outward.

Our framework is also different compared with the spherical Bondi
accretion\cite{bondi52}, where a sound speed of the gas is
considered, and a numerical correction factor
$\alpha_{\mathrm{B}}\approx100$ (see Ref.~\refcite{springel05}),
which depends on the mass profile and gas equation of state, is
inserted in Eq. (\ref{Eq_02}). Some authors, by analyzing the role
of angular momentum, refer to that special case of spherical Bondi
accretion, finding that it fails systematically to reproduce their
numerical simulations\cite{hopkins10}. On the other hand, other
authors, by using Chandra X-ray observations of several nearby
elliptical galaxies, observed a tight correlation between the
Bondi accretion rates (calculated from the observed gas
temperature and density profiles as well as from the estimated
black hole masses) and the power emerging from these systems in
relativistic jets\cite{allen06}. They concluded that the Bondi
formulae provide a reasonable description of the accretion process
in these systems, despite the likely presence of angular momentum
in the accreting gas.

We can study the effect of making the BHL theory a real assumption
of our model. Let us analyze a different case (based on different
hypotheses) from the one
considered in section 2, assuming that the angular momentum is
very small so the BHL theory is not only a fair approximation, but
it works very well. Actually we can suppose that the photon is
emitted by the black hole with a velocity $\vec{c}$ in the same
direction of the arriving gas particle on the accretion line. It
means that $\vec{c}$  forms an angle $\alpha$ with this line and
with the radial component of gas velocity $V_{\mathrm{in}}$ and
hence $A = V_{\mathrm{rot}}/V_{\mathrm{in}} = \tan(\alpha)$. In
this case the conservation of angular momentum can be written
\begin{equation}
M_{\mathrm{acc}} \, R_{\mathrm{e}} \, V_{\mathrm{rot}} =
M_{\mathrm{acc}} \, R_{\mathrm{e}} \, V_{\mathrm{in}} \tan(\alpha)
= c \, M_{\mathrm{rad}} \, R_{\mathrm{A}} \sin(\alpha).%
\label{Eq_18}
\end{equation}
If we suppose that the accretion occurs mainly for particles with
low angular momentum  ($A <<1$), we can simplify the above
equation because for $\alpha \simeq 0$ we have $\tan(\alpha)
\simeq \sin(\alpha)$. In this particular case, our fundamental Eq.
(4) becomes
\begin{equation}
M_{\bullet} =  \frac{R_{\mathrm{e}} \sigma^3}{2 \, \varepsilon \, G \, c }
\label{Eq_19}
\end{equation}
and the effect to consider  the BHL theory as a real hypothesis of
the model reduces to the disappearing of the factor $A$ from the
fundamental Eq. (4), and all the calculations of the first part of
the paper are still valid considering $\hat{\varepsilon} = 2
\varepsilon$. We reported again the data of the Hu
sample\cite{hu09} in Figure 2 together with the lines representing
the values of the efficiency coefficient $\varepsilon$ predicted
by Eq. (19). So, {\it given the SMBH mass and the effective radius and dispersion velocity of the
host galaxy, the equation (19) allows a quick estimate of the efficiency
of a black hole}. In this way the resulting diagram (Fig. 2)
can be used to classify the black holes in terms of their efficiency provided that the host
galaxies satisfy the hypotheses of the model. For example,
 in our sample all the galaxies have a predicted $\varepsilon
<0.25$, but the prediction could be not valid, as expected, for some pseudobulges for
which the approximation of the model $A <<1$ does not work.
\begin{figure}[pb]
\centerline{\psfig{file=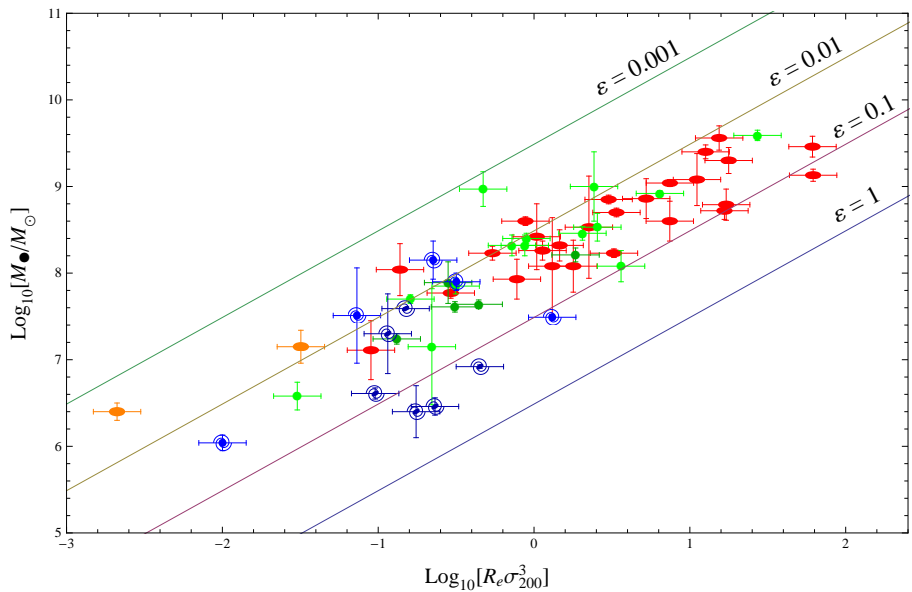,width=12.cm}} %
\vspace*{8pt} \caption{The data of the considered sample in the
$M_{\bullet}-R_{\mathrm{e}}\sigma_{200}^3$ plane. The symbols are
the same of Figure 1. Lines of constant values of the efficiency
coefficient $\varepsilon$ are shown, see Eq. (\ref{Eq_19}). All
the galaxies have $\varepsilon < 0.25$.}
\end{figure}

\section*{Appendix B}
The slopes of the $M_{\bullet} - M_\mathrm{G}$ and $M_{\bullet} -
M_\mathrm{G} \sigma^2$ relations remain very stable  changing the
sample of data or the fitting method. In Ref 28 we have used one
fitting method and three different samples and the resulting
slopes were (1.18, 1.22, 1.27) for the $M_{\bullet} -
M_\mathrm{G}$ relation  and (0.83, 0.86, 0.91) for the
$M_{\bullet} - M_\mathrm{G} \sigma^2$ relation. Viceversa, in Ref.
27 we have used only one sample and three different fitting
procedures and the  slopes are still stable with values (1.15,
0.98, 1.07) and  (0.80, 0.74, 0.78) respectively. These values are
in agreement with the results of other authors. For instance,
Hopkins et al.\cite{hopkins07b} find 1.05 for $M_{\bullet} -
M_\mathrm{G}$ and 0.71 for $M_{\bullet} - M_\mathrm{G} \sigma^2$,
whereas Soker et al.\cite{soker10} find 1.07 and 0.74
respectively.

The $M_{\bullet} - \sigma$ relation has a different behavior since
its slope strongly depends on the fitting method (5.06, 4.46,
4.25) in Ref. \refcite{feoli09}, and on different samples (5.26,
4.99, 5.83) in Ref. \refcite{feoli10}. In this case Hopkins et
al.\cite{hopkins07b} find 3.96, Soker\cite{soker10} 4.18,
Ferrarese and Ford\cite{Ferford05} 4.86 and 5.1 and Hu\cite{Hu08}
a slope that changes from 4.01 to 5.62 depending on the considered
subsample of data.

The oscillation of the slope between 3.9 and 5.9 could be due in
part to the fitting method (FITEXY finds the first minimum of
$\chi^2$ and gives a slope; if one introduces the intrinsic
scatter and refits until the reduced $\chi^2 = 1$, a smaller slope
is generally obtained\cite{tremaine02}) and in part to the
presence of pseudobulges in the sample. In their last paper Sani
et. al.\cite{sani11} find the same value (4.00) for the slope
using three different methods, but selecting only the classical
bulges. By introducing also the pseudobulges in their sample, the
slope increases and the result becomes no longer stable.

\end{document}